\documentclass[12pt]{amsart}
\usepackage{amsmath}
\usepackage{amssymb}
\usepackage{bbm}
\usepackage{graphicx}
\usepackage{fullpage}
\newtheorem{prop}{Proposition}

\newtheorem{thm}{Theorem}
\newtheorem{rem}{Remark}
\newtheorem{lem}{Lemma}
\newtheorem{ex}{Example}

\newtheorem{Def}{Definition}

\def\x{{\mathbf x}}
\def\y{{\mathbf y}}
\def\q{{\mathbf q}}

\DeclareMathOperator*{\argmin}{arg\,min}

\begin{document}
\pagestyle{plain}
\newenvironment{frcseries}{\fontfamily{frc} \selectfont}{}
\newcommand{\textfrc}[1]{{\frcseries #1}}
\newcommand{\mathfrc}[1]{\text{\textfrc{#1}}}

\newcommand{\sd}{\Sigma\Delta}
\newcommand{\R}{\mathbbm{R}}
\newcommand{\A}{\mathcal{A}}
\newcommand{\Q}{\mathcal{Q}}
% Title.
% ------
\title{Near-Optimal Encoding for  Sigma-Delta Quantization of Finite Frame Expansions}
%
% Single address.
% ---------------
\author{Mark Iwen}
\address{Mark Iwen \newline \indent Department of Mathematics, Michigan State University \newline \indent Department of Electrical and Computer Engineering, Michigan State University}
\email{iwenmark@msu.edu}
\thanks{M.I. was supported in part by NSA grant H98230-13-1-0275. R.S. was supported in part by a Banting Postdoctoral Fellowship, administered by the Natural Sciences and Engineering Research Council of Canada (NSERC). The majority of the work reported on herein was completed while the authors were visiting assistant professors at Duke University.}
\author{Rayan Saab}
\address{Rayan Saab \newline \indent Department of Mathematics, University of California, San Diego}
\email{rsaab@ucsd.edu}

\begin{abstract}
In this paper we investigate encoding the bit-stream resulting from coarse Sigma-Delta quantization of finite frame  expansions (i.e., overdetermined representations) of vectors. We show that for a wide range of finite-frames, including random frames and piecewise smooth frames, there exists a simple encoding algorithm ---acting only on the Sigma-Delta bit stream--- and an associated decoding algorithm that together yield an approximation error which decays exponentially in the number of bits used.  The encoding strategy consists of applying a discrete random operator to the Sigma-Delta bit stream and assigning a binary codeword to the result. The reconstruction procedure is essentially linear and equivalent to solving a least squares minimization problem.\\
 

\end{abstract}

\maketitle
\thispagestyle{empty}

%%%%%%%
%%%%%%%
\section{Introduction}
\label{sec:Intro}
In the modern era, the first step in signal processing consists of obtaining a digital representation of the signal of interest, i.e., quantizing it. This enables one to  store, transmit, process, and analyze the signal via digital devices. Sigma-Delta ($\sd$) quantization was proposed in the 1960's as a quantization scheme for digitizing band-limited signals (see, e.g., \cite{inose1963unity}). Since then, and especially with the advent of very large scale integration (VLSI) technology, $\sd$ schemes have seen extensive use in the engineering community for analog-to-digital conversion of, for example, audio signals (cf. \cite{norsworthy1997delta}). In the mathematical community, $\sd$ quantization has seen increasing interest since the work of Daubechies and Devore \cite{daub-dev}. In this paper, we are interested in efficiently encoding the bit-stream resulting from $\sd$ quantization of finite-frame expansions. Here one models the signal as an element in a finite dimensional space, and its samples as inner products with a spanning set of vectors. The goal is, using only the samples, to obtain a digital representation of the original signal that allows its high fidelity reconstruction. 

\subsection{Overview and prior work} For concreteness, let the vectors $\{\mathbf{f}_i\}_{i=1}^N\subset \R^d$  form a finite frame for $\R^d$. In other words, suppose there exist constants $0<A\leq B<\infty$ such that the {\em frame matrix}
$F\in \R^{N\times d}$ (with the vectors $\mathbf{f}_i$ as its rows) satisfies
$$A\|\x\|_2^2\leq \|F\x\|_2^2 \leq B\|\x\|_2^2$$
for all $x\in \R^d$. Thus any full rank matrix is a frame matrix. In the context of data acquisition, finite frames are useful at modeling the sampling (i.e., measurement) process. In various applications, the measurement vector can be expressed as \begin{equation}\y:=F\x \in \R^N \label{eq:measurements} .\end{equation} For example, in imaging applications, ``multiplex" systems (see, e.g., \cite{brady2002multiplex}) collect linear combinations of the pixels of interest, thus their measurement vectors can the be represented using \eqref{eq:measurements}. Such systems have been devised using coded apertures (see, e.g., \cite{kohman1989coded}), as well as  digital micro-mirror arrays (e.g., \cite{duarte2008single}). Indeed, by simply collecting more measurements than the ambient dimension of the image, as is often the case, and ensuring that $F$ is full rank, we find ourselves in the finite frame setting. Similarly, systems that acquire finite dimensional signals using filter banks allow their measurement process to be modeled via \eqref{eq:measurements}. For a more in-depth treatment of finite frames and filter banks, see \cite{fickus2013finite}. 

In order to allow digital storage and computer processing, including the recovery of $x$, one must quantize the finite frame expansion \eqref{eq:measurements}. Quantizing the finite frame expansion of $\x$ consists of replacing the entries of the measurement vector $\y:=F\x \in \R^N$ with elements from a finite set. Giving these elements binary labels then enables digital
storage and transmission of the quantized measurements. To be precise, let $\A\subset \R$ be a finite set (the quantization alphabet), and let
$\mathcal{X}$ be a compact set in $\R^d$.  A quantization
scheme is a map $${\Q}: F\mathcal{X} \mapsto \mathcal{A}^N$$ and a
reconstruction scheme is an ``inverse" map $$\Delta: \mathcal{A}^N
\mapsto \R^d.$$ Note that, depending on the practical application, one may require that the quantization schemes satisfy certain properties. For example, it is often preferred that the quantization scheme acts progressively on the measurements, i.e., as they arrive, to avoid storing too many analog quantities. Nevertheless, one seeks quantization and reconstruction schemes with approximation errors $\| x-
\Delta(\Q(Fx))\|_2$ that are as small as possible for all $x\in
\mathcal{X}$. 

$\sd$ quantization schemes form an important class of such progressive quantizers, and there has been a body of research focusing on their application to the finite frame setting. In particular, the research on $\sd$ quantization of finite frame expansions has focused on the decay of the approximation error as a function of the number of measurements and has typically considered $\mathcal{X}=\mathcal{B}^d$, the Euclidean ball in $\R^d$. The work of Benedetto, Powell, and Y{\i}lmaz \cite{benedetto2006sigma} first showed that the reconstruction error associated with $1$st order $\sd$ quantization decays linearly in the number of measurements. Several results followed, improving on this linear error decay by using various combinations of specialized frames, higher order quantization schemes, and different reconstruction techniques. Blum et al. \cite{blum:sdf} showed that frames with certain smoothness properties allow for polynomial decay in the $\sd$ reconstruction error, provided appropriate alternative dual frames are used for reconstruction. Motivated by applications in compressed sensing, a similar result \cite{GLPSY12} was shown for random frames whose elements are Gaussian random variables.  This was followed by the results of \cite{krahmer2012root} and \cite{KSY12} showing that there exist (deterministic and random) frames for which higher order $\sd$ schemes yield approximation errors that behave like $e^{-c\sqrt{\frac{N}{d}}}$, where $c$ is a constant. Specifically, in \cite{krahmer2012root, KSY12} this root-exponential accuracy is achieved by carefully  choosing the order of the scheme as a function of the oversampling rate $N/d$. For a more comprehensive review of $\sd$ schemes applied to finite frames, see \cite{PSY12}.

While the above results progressively improved on the {\em coding efficiency} of $\sd$ quantization, it remains true that even the root-exponential performance $e^{-c\sqrt{\frac{N}{d}}}$ of \cite{krahmer2012root,KSY12} is generally sub-optimal from an information theoretic perspective, including in the case where $\mathcal{X}=\mathcal{B}^d$. %To be more precise, it is well known that to cover $\mathcal{B}^d$ with balls of radius $\epsilon$, one needs at least $\left(\frac{c}{\epsilon}\right)^d$  such $\epsilon$-balls, and there exists a covering with no more than $\left(\frac{c'}{\epsilon}\right)^d$ elements ---here $c, c'$ are constants. {\bf MARK: I can prove this with $c=1/2$ and $c'=3$ using $\epsilon-$nets and $\epsilon$-distinguishable  sets (equivalently covering and packing numbers) but I can't easily find a reference, should we expend half a page in the appendix on it, or do we just leave it as folklore like this?}. Thus, quantizing $\mathcal{B}^d$ via an optimal map requires $d\ln{\frac{c}{\epsilon}}$ bits, or, viewed slightly differently: optimally quantizing $\mathcal{B}^d$ with $b$-bits yields an approximation error of the form $e^{-c \frac{b}{d}}$. Observing that the number of bits that result from a $\sd$ scheme is proportional to $N$ and that the best known error rates are root-exponential in $N$, we conclude that $\sd$ schemes are sub-optimal. 
To be more precise, any quantization scheme tasked with encoding all possible points in $\mathcal{B}^d$ to within $\epsilon$-accuracy must produce outputs which, in the case of an optimal encoder, each correspond to a unique subset of the unit ball having radius at most $\epsilon$.  Covering each of these subsets with a ball of radius at most $\epsilon$ then produces an $\epsilon$-cover of $\mathcal{B}^d$.  A simple volume argument now shows that covering $\mathcal{B}^d$ with balls of radius $\epsilon$ requires one to use at least $\left(\frac{1}{\epsilon}\right)^d$ such $\epsilon$-balls.\footnote{Moreover, there exists a covering with no more than $\left(\frac{3}{\epsilon}\right)^d$ elements (see, e.g., \cite{lorentz1996constructive}).}  Thus, quantizing $\mathcal{B}^d$ via an optimal map requires at least $d\ln{\frac{c}{\epsilon}}$ bits, or, viewed slightly differently: optimally quantizing $\mathcal{B}^d$ with $b$-bits yields an approximation error of the form $e^{-c \frac{b}{d}}$, where $c \in \mathbbm{R}^+$ is a universal constant. Observing that the number of bits that result from a $\sd$ scheme is proportional to $N$ and that the best known error rates are root-exponential in $N$, we conclude that $\sd$ schemes are sub-optimal. 

This fact has been recognized in the mathematical literature on $\sd$ quantization. In particular, in the case where $d=1$ and the frame $F$ is the $N\times 1$ repetition frame with $F_{i1}=1$ for all $i \in [N]$, there has been research seeking upper bounds on the maximum number of possible $\sd$ bit-streams (cf. \cite{hein1992new}, \cite{gunturk2001robustness}, \cite{ayaz2009sigma}). For example, \cite{hein1992new} showed that asymptotically in $N$, the number of bit-streams is bounded by $O(N^2)$ for first order single-bit $\sd$ schemes with certain initial conditions. This indicates that by losslessly encoding the possible $\sd$ bitstreams into codewords of length $O(\log(N))$, one can achieve the desired exponential error rates. However, to do that one needs to identify the $O(N^2)$ achievable sequences from among the $2^N$ potential ones, which to our knowledge is an unsolved problem. Moreover, to our knowledge, not much is known about the number of codewords generated by $\sd$ quantization in more general settings. 

To help remedy this situation, this paper introduces a potentially {\em{lossy encoding}} stage, consisting of the map $$\mathcal{E}: \mathcal{A}^N \mapsto {\mathcal{C}},$$ where $\mathcal{C}$ is such that $|\mathcal{C}| \ll |\mathcal{A}^N| $. Consequently, $\log_2|\mathcal{C}|$ bits are sufficient for digitally representing the output of this encoder. To accommodate this additional encoding, the reconstruction is modified to approximate $\x$ directly from $\mathcal{C}$. Thus, we propose a decoder
$$\Delta: \mathcal{C} \mapsto \R^d,$$ 
where both the proposed decoder, $\Delta$, and the proposed encoding map, $\mathcal{E}$, are \textit{linear}, hence computationally efficient.

\subsection{Contributions}For stable $\sd$ quantization schemes, % our results can be summarized as follows. 
%\vspace{+5pt}
%
%\begin{enumerate}\itemsep +5pt
%\item[\bf 1.]
 we show that there exists an encoding scheme $\mathcal{E}$ acting on the output $\mathcal{Q}(F\x)$ of the quantization, and a decoding scheme $\Delta$, such that
\begin{equation}
 \left.  \begin{array} {lll}
 \epsilon_{\sd}&:=&\max\limits_{x\in\mathcal{B}^d} \Big \| x- \Delta \Big(   \mathcal{E}\big(\Q(Fx) \big)   \Big) \Big \|_2 \leq CN^{-\alpha}
\\ b_{\sd}&:=&\ln|\mathcal{C}| \leq C' d\ln N, 
\end{array} \right\}%
\implies \epsilon_{\sd} \leq \exp\left({-{c} \frac{b_{\sd}}{d}}\right). 
\nonumber\end{equation}
where $\alpha$, $C$, $C'$, and $c$ are positive constants that depend on the $\sd$ scheme and $d$.  More specifically: 
\begin{itemize}
\item[\bf 1.] We show that  there exist frames (the Sobolev self-dual frames), for which encoding by random subsampling of the integrated $\sd$ bit-stream (and labeling the output) yields  an {\em essentially optimal rate-distortion tradeoff up to logarithmic factors of $d$}. 
\item[\bf 2.] We show that random \emph{Bernoulli} matrices in $\R^{m\times d}$, with $m\approx d$, are \emph{universal} encoders. Provided one has a good frame for $\sd$ quantization, such Bernoulli matrices  yield an {\em optimal rate-distortion tradeoff, up to constants}. 
\item[\bf 3.] We show that in both cases above, the decoding can be done linearly and we provide an explicit expression for the decoder. 
\end{itemize}
%\end{enumerate}
These contributions are made explicit in Theorems \ref{thm:ResExp} and \ref{thm:BerExp}. Additionally, we note that $\sd$ schemes (see Section \ref{sec:Prelim}) act progressively on the samples $\y=F\x$, and do not require explicit knowledge of the frame that produced $\y$. Similarly, the Bernoulli encoding of the $\sd$ bit-stream does not require knowledge of the underlying frame. Nevertheless, and somewhat surprisingly, this encoding method allows the compression of the $\sd$-bitstream in a near optimal manner. It also allows the decoding to be done linearly via an operator $\R^m \mapsto \R^d$, hence in time proportional to $md$, as opposed to time proportional to $Nd$ needed (in general) for reconstructing a signal from its unencoded $\sd$ bitstream. One of the favorable properties of coarse $\sd$ quantization schemes is their robustness to certain errors that can arise in practice due to (for example) circuit imperfections (cf. \cite{daub-dev}). Such imperfections can affect the elements that implement scalar quantization (i.e., assigning discrete values to continuous ones by toggling at a threshold), or multiplication. We remark that our methods for compressing the $\sd$ bit-stream inherit whatever robustness properties the original $\sd$ quantizer possesses. In other words, by compressing the bit-stream, we do not lose any of the desirable properties of $\sd$ quantization. 

\subsection{Organization} In Section \ref{sec:Prelim}, we introduce notation and provide a mathematical overview of $\sd$ quantization. We also state certain results on random matrices, in particular Johnson-Lindenstrauss embeddings, which will be useful in the remainder of the paper. %In Sections \ref{sec:randSamp} and \ref{sec:BernSamp} we present and prove our main results.
In Section \ref{sec:randSamp} we show that random subsampling of the discretely integrated $\Sigma\Delta$ bit-stream allows a linear decoder to achieve exponentially decaying reconstruction error, uniformly for all $x\in\mathcal{B}^d$. This result pertains to a particular choice of frames, the Sobolev self-dual frames \cite{krahmer2012root}, and is contingent on using $1$st order $\sd$ schemes. %While near optimal, our result in this section contains an extra logarithmic dependency on $d$, compared to an optimal quantizer. 
In Section \ref{sec:BernSamp} we instead use a Bernoulli matrix for reducing the dimensionality of the integrated $\sd$-bit stream. Here, our result is more general and applies to stable $\sd$ schemes of arbitrary order, as well as to a large family of smooth and random frames. Finally, in Section \ref{sec:Numerical} we illustrate our results with numerical experiments. 

%%%%%%%
%%%%%%%
\section{Preliminaries}
\label{sec:Prelim}

Below we will denote the set $\{ 1, 2, \dots, n-1, n \} \subset \mathbbm{N}$ by $[n]$.  For any matrix $M \in \mathbbm{R}^{m \times N}$ we will denote the $j^{\rm th}$ column of $M$ by ${\bf M}_j \in \mathbbm{R}^{m}$.  Furthermore, for a given subset $\mathcal{S} = \{ s_1, \dots, s_n \} \subset [N]$ with $s_1 < s_2 < \dots < s_n$, we will let $M_{\mathcal{S}} \in \mathbbm{R}^{m \times n}$ denote the submatrix of $M$ given by
$$M_{\mathcal{S}} := \left( {\bf M}_{s_1} \dots {\bf M}_{s_n} \right).$$
The transpose of a matrix, $M \in \mathbbm{R}^{m \times N}$, will be denoted by $M^{\rm T} \in \mathbbm{R}^{N \times m}$, and the singular values of any matrix $M \in \mathbbm{R}^{m \times N}$ will always be ordered  as $\sigma_1(M) \geq \sigma_2(M) \geq \dots \geq \sigma_{\min(m,N)}(M) \geq 0.$  We will denote the standard indicator function by
$$\delta_{i,j} := \left\{ \begin{array}{ll} 1 & \textrm{if}~i=j\\ 0 & \textrm{if}~i \neq j\end{array} \right.,$$
for $i,j \in \mathbbm{N}$.
%We will refer to any full rank matrix $F \in \mathbbm{R}^{N \times d}$, with $N \geq d$, as a \textit{frame matrix}.  Given a frame matrix $F$, we will refer to $A,B \in \mathbbm{R}^+$ as the the \textit{frame bounds} for $F$ if $A$ and $B$ are the largest and smallest values, respectively, such that $$A \| {\bf x} \|^2_2 ~\leq~ \left \| F {\bf x} \right\|^2_2 ~\leq~B \| {\bf x} \|^2_2$$ holds for all ${\bf x} \in \mathbbm{R}^d$.  If $A = B$ we will say that the frame $F$ is a \textit{tight frame}.  
Finally, given a frame matrix $F$, we define its Moore-Penrose pseudo-inverse to be %$F^\dagger ~:=~V^{\rm T} \Sigma^{-1} U^{\rm T}$, where $F = U \Sigma V$ is the singular value decomposition of $F$.
$F^\dagger~:=({F^TF})^{-1}F^T.$

\subsection{Sigma-Delta Quantization}
\label{sec:SigmaDelt}

Let $\mathcal{B}^d$ be the Euclidean unit ball in $\mathbbm{R}^d$.  Given ${\bf x} \in \mathcal{B}^d$, and a frame matrix $F \in \mathbbm{R}^{N \times d}$, %we will consider Sigma-Delta$\left( \Sigma \Delta \right)$ quantization schemes which quantize the frame expansion of ${\bf x}$ in terms of $F$ as some ${\bf q} \in \mathbbm{Z}^N$.  The simplest such quantization scheme considered herein, in Section~\ref{sec:randSamp} below, is the \textit{first order scheme} which computes a vector ${\bf q} \in \{ -1, 1 \}^N$ for each ${\bf x} \in \mathcal{B}^d$ via the following system of equations:  
the simplest $\sd$ quantization scheme considered herein, in Section~\ref{sec:randSamp}, is the \textit{single bit first order  greedy scheme}. Given $\y=F\x$,  this scheme computes a vector ${\bf q} \in \{ -1, 1 \}^N$ via the following recursion with initial condition $u_0=0$:  
%\begin{equation}
%{\bf y} = F {\bf x},~~
%u_0 = 0,
%\label{equ:FirstOrdInit}
%\end{equation}
\begin{equation}
q_i = \textrm{sign} \left( y_i + u_{i-1} \right), %\textrm{and}
\label{equ:FirstOrdq}
\end{equation}
\begin{equation}
u_i =  y_i + u_{i-1} - q_i
\label{equ:FirstOrdu}
\end{equation}
for all $i \in [N]$.  To analyze this scheme as well as higher order schemes, it will be convenient to introduce the \textit{difference matrix}, $D \in \mathbbm{R}^{N \times N}$, given by
\begin{equation}
D_{i,j} := \left\{ \begin{array}{ll} 1 & \textrm{if}~i=j\\ -1 & \textrm{if}~i = j+1\\ 0 & \textrm{otherwise}\end{array} \right..
\label{Def:diffMat}
\end{equation}
We may restate the relationships between ${\bf x}$, ${\bf u}$, and ${\bf q}$ resulting from the above scheme as%$\eqref{equ:FirstOrdInit} - \eqref{equ:FirstOrdu}$ as
\begin{equation}
D {\bf u} ~=~ F {\bf x} - {\bf q}.
\label{equ:LinSigDelt}
\end{equation}
Furthermore, a short induction argument shows that $|u_i| \leq 1$ for all $i \in [N]$ provided that $|y_i| \leq 1$ for all $i \in [N]$.

More generally, for a given alphabet $\A$ and $r \in \mathbbm{Z}^{+}$ we may employ an \textit{$r^{\rm th}$-order $\Sigma \Delta$ quantization scheme}  with quantization rule $\rho:  \mathbbm{R}^{r+1} \mapsto \mathbbm{R}$ and scalar quantizer $Q:\mathbbm{R}\mapsto \mathcal{A}$. 
%which quantizes a given ${\bf x} \in \mathcal{B}^d$ in terms of a frame matrix $F$ as some ${\bf q} \in \mathcal{A}^N \subset \mathbbm{Z}^N$.  Here the set $\mathcal{A} \subset \mathbbm{Z}$ is called the \textit{quantization alphabet}.  A choice of the quantization alphabet, together with an associated surjective \textit{quantizing function} 
%$$Q:  \mathbbm{R}^{r+1} \rightarrow \mathcal{A},$$ 
%determines the $r^{\rm th}$-order $\Sigma \Delta$ schemes considered herein via the following system of equations:
Such a scheme, with initial conditions $u_0 =u_{-1}=\cdots= u_{1-r}=0$, computes $\q\in \A^N$ via the recursion 
%\begin{equation}
%{\bf y} = F {\bf x},~~~~~u_0 =u_{-1}=\cdots= u_{1-r}=0,
%\label{equ:rthOrdInit}
%\end{equation}
\begin{equation}
q_i = Q\left(\rho(y_i,u_{i-1},u_{i-2},\dots,u_{i-r})\right),% \textrm{and}
\label{equ:rthOrdq}
\end{equation}
\begin{equation}
u_i =  y_i - q_i - \sum^{r}_{j=1} {r \choose j} (-1)^j u_{i-j}
\label{equ:rthOrdu}
\end{equation}
for all $i \in [N]$. Here, the scalar quantizer $Q$ is defined via its action $$Q(v) = \argmin\limits_{q\in\A}|q-v|.$$   In this paper, we focus on {\em midrise} alphabets of the form 
\begin{equation}\label{eq:midrise}\mathcal{A}^{\delta}_K=\{\pm(2n-1)\delta/2: n \in [K]\},\end{equation} where $\delta$ denotes the quantization step size.  For example, when $K=1$, we have the $1$-bit alphabet $\mathcal{A}_1^{\delta}=\{\pm \frac{\delta}{2} \}$. As before, we may restate the relationships between ${\bf x}$, ${\bf u}$, and ${\bf q}$ as %resulting from $\eqref{equ:rthOrdInit} - \eqref{equ:rthOrdu}$ as
\begin{equation}
D^r {\bf u} ~=~ F {\bf x} - {\bf q}.
\label{equ:LinrSigDelt}
\end{equation}
As in the case of the first order scheme, we will ultimately need a bound on $\| {\bf u} \|_{\infty} := \max_{i \in [N]} |u_i|$ below.  Hence, we restrict our attention to  \textit{stable $r^{\rm th}$-order schemes}.  That is, $r^{th}$-order schemes for which \eqref{equ:rthOrdq} and \eqref{equ:rthOrdu} are guaranteed to always produce vectors ${\bf u} \in \mathbbm{R}^N$ having $\| {\bf u} \|_{\infty} \leq {C}_{\rho,Q}(r)$ for all $N \in \mathbbm{N}$, and ${\bf y} \in \mathbbm{R}^N$ with $\|{\bf y}\|_\infty \leq 1$. Moreover, for our definition of stability we require that $C_{\rho,Q}:  \mathbbm{N} \mapsto \mathbbm{R}^+$ be entirely independent of both $N$ and ${\bf y}$.  Finally, it is important to note that stable $r^{\rm th}$-order $\Sigma \Delta$ schemes with $C_{\rho,Q}(r) = O(r^r)$ do indeed exist (see, e.g., \cite{gunturk2003one,deift2011optimal}), even when $\mathcal{A}$ is a $1$-bit alphabet. In particular, we cite the following proposition (see  \cite{krahmer2012root}).

\begin{prop}\label{prop:best_known}
There exists a universal constant $c>0$ such that for any midrise quantization alphabet $\mathcal{A}=\mathcal{A}^\delta_L$, for any order $r\in\mathbbm{N}$, and for all $\eta<\delta\left(L-\frac{1}{2}\right)$, there exists an $r$th order $\Sigma\Delta$ scheme which is stable for all input signals $y$ with $\|y\|_\infty\leq \eta$.  It has
\begin{equation}\label{eq:best_known}
\|u\|_\infty \leq c C^r r^r \frac{\delta}{2},
\end{equation}
where $C=\left(\left\lceil \frac{\pi^2}{(\cosh^{-1} \gamma)^2} \right\rceil \frac{e}{\pi}\right)$ and $\gamma:=2L-\frac{2\eta}{\delta}$.
\end{prop}

In what follows we will need the singular value decomposition of $D$ essentially computed by von Neumann in \cite{vonNeumann1941} (see also \cite{krahmer2012root}).  It is $D ~=~ U \Sigma V^T$, where 
\begin{equation}
U_{i,j}~=~\sqrt{\frac{2}{N + 1/2}} \cos \left( \frac{2(i-1/2)(N - j + 1/2) \pi}{2N + 1} \right),
\label{equ:LsingVecsD}
\end{equation}
\begin{equation}
\Sigma_{i,j} ~=~ \delta_{i,j}\sigma_j(D) ~=~ 2 \delta_{i,j} \cos \left( \frac{j \pi}{2N + 1} \right),
\label{equ:SingValsD}
\end{equation}
and
\begin{equation}
V_{i,j} ~=~ (-1)^{j+1} \sqrt{\frac{2}{N + 1/2}} \sin \left( \frac{2 i j}{2N + 1} \pi \right).
\label{equ:RsingVecsD}
\end{equation}
Note that the difference matrix, $D$, is full rank (e.g., see~\eqref{equ:SingValsD}).  Thus, we may rearrange \eqref{equ:LinSigDelt} to obtain
\begin{equation}
{\bf u} ~=~ D^{-1}F{\bf x} - D^{-1}{\bf q}.
\label{equ:Recon}
\end{equation}
More generally, rearranging \eqref{equ:LinrSigDelt} tells us that
\begin{equation}
{\bf u} ~=~ D^{-r}F{\bf x} - D^{-r}{\bf q}
\label{equ:ReconOrdr}
\end{equation}
for any $r^{\rm th}$-order scheme.

\subsection{Johnson-Lindenstrauss Embeddings %, Restricted Isometries, 
 and Bounded Orthonormal Systems}

We will utilize  \textit{linear Johnson-Lindenstrauss embeddings} \cite{JLoriginal,frankl1988johnson,achlioptas2001,dasgupta2003elementary,baraniuk2008simple,krahmer2010new} of a given finite set $\mathcal{S} \subset \mathbbm{R}^N$ into $\mathbbm{R}^m$.%, as well as the closely related \textit{Restricted Isometry Property} \cite{candes2005decoding,baraniuk2008simple,HolgerBook}, below.  These are defined as follows:
\begin{Def}
Let $\epsilon, p \in (0, 1)$, and $\mathcal{S} \subset \mathbbm{R}^N$ be finite.  An $m \times N$ matrix $M$ is a linear Johnson-Lindenstrauss embedding of $\mathcal{S}$ into $\mathbbm{R}^m$ if
the following holds with probability at least $1-p$:
$$(1 - \epsilon) \| ~{\bf u} - {\bf v}~ \|_2^2 \leq \| ~M{\bf u} - M{\bf v}~ \|_2^2 \leq (1 + \epsilon) \| ~{\bf u} - {\bf v}~ \|_2^2$$
for all ${\bf u},{\bf v} \in \mathcal{S}$.  In this case we will say that $M$ is a JL($N$,$m$,$\epsilon$,$p$)-embedding of $\mathcal{S}$ into $\mathbbm{R}^m$.
\label{Def:JL}
\end{Def}  
%\begin{Def}
%Let $N, d \in \mathbbm{N}$, and $\epsilon,p \in (0, 1)$. An $m \times N$ matrix $M$ has the Restricted Isometry Property if the following holds with probability at least $1-p$:
%\begin{equation}
%(1 - \epsilon) \| ~{\bf x}~\|_2^2 \leq \| ~M {\bf x}~\|_2^2 \leq (1 + \epsilon) \| ~{\bf x}~ \|_2^2
%\label{equ:RIP}
%\end{equation}
%for all ${\bf x} \in \mathbbm{R}^N$ containing at most $d$ nonzero coordinates.  In this case we will say that $M$ is RIP($N$,$d$,$\epsilon$,$p$). 
%\end{Def}

We will say that a matrix $B \in \{-1,1 \}^{m \times N}$ is a \textit{Bernoulli random matrix} iff each of its entries is independently and identically distributed so that
$$\mathbbm{P} \left[B_{i,j} = 1 \right] ~=~ \mathbbm{P} \left[B_{i,j} = -1 \right]~=~\frac{1}{2}$$
for all $i \in [m]$ and $j \in [N]$.  The following theorem is proven in \cite{achlioptas2001}. %and \cite{BernoulliRIP,HolgerBook}, respectively.%\footnote{The specific form of the lower bound used for $m$ in Theorem~\ref{thm:HolgB} is taken from Theorem 9.8 of \cite{HolgerBook}.}
\begin{thm}
Let $m, N \in \mathbbm{N}$, $\mathcal{S} \subset \mathbbm{R}^N$ finite, and $\epsilon, p \in (0,1)$.  Let $B \in \{ -1, 1\}^{m \times N}$ be a Bernoulli random matrix, and set $\widetilde{B} = \frac{1}{\sqrt{m}}B$.  Then, $\widetilde{B}$ will be a JL($N$,$m$,$\epsilon$,$p$)-embedding of $\mathcal{S}$ into $\mathbbm{R}^m$ provided that $m \geq \frac{4 + 2 \log_{|\mathcal{S}|}(1/p)}{\epsilon^2/2 - \epsilon^3/3} \ln |\mathcal{S}|$.
\label{thm:BernJL}
\end{thm}
%\begin{thm}
%There exists a fixed universal constant, $C > 0$, such that the following holds for all $d, m, N \in \mathbbm{N}$ and $\epsilon, p \in (0,1)$:   Let $\mathcal{S} \subset [N]$ have have cardinality $|\mathcal{S}| = d$.  Draw an $m \times N$ Bernoulli random matrix, $B$, and set $\widetilde{B} = \frac{1}{\sqrt{m}}B$.  If $m \geq C \epsilon^{-2} \left( 7 d + 2 \ln \left(2/ p \right)\right)$ then we will have
%$$\sqrt{1 - \epsilon} ~\leq~ \sigma_d \left(\widetilde{B}_{\mathcal{S}} \right)~\leq~\sigma_1\left( \widetilde{B}_{\mathcal{S}} \right) ~\leq~ \sqrt{1 + \epsilon}$$
%with probability at least $1-p$.
%\label{thm:HolgB}
%\end{thm}

Let $\mathcal{D} \subset \mathbbm{R}^n$ be endowed with a probability measure $\mu$.  Further, let $\Psi = \{ \psi_1, \dots, \psi_N \}$ be an orthonormal set of real-valued functions on $\mathcal{D}$ so that
$$\int_{\mathcal{D}} \psi_i\left(~{\bf t} ~\right) \overline{\psi_j \left(~{\bf t}~ \right)} d \mu \left( {\bf t} \right) ~=~ \delta_{i,j}.$$
We will refer to any such $\Psi$ as an \textit{orthonormal system}.  More specifically, we utilize a particular type of orthonormal system:
\begin{Def}
We call $\Psi = \{ \psi_1, \dots, \psi_N \}$ a bounded orthonormal system with constant $K \in \mathbbm{R}^+$ if 
$$\left\| \psi_k \right\|_{\infty} := \sup_{{\bf t} ~\in~ \mathcal{D}} \left| \psi \left(~ {\bf t}~\right) \right| ~\leq~K~~\textrm{for all}~k \in [N].$$
\label{Def:BOS}
\end{Def}

For any orthonormal system, $\Psi$, on $\mathcal{D} \subset \mathbbm{R}^n$ with probability measure $\mu$, we may create an associated \textit{random sampling matrix}, $R' \in \mathbbm{R}^{m \times N}$, as follows:  First, select $m$ points ${\bf t}_1, \dots, {\bf t}_m \in \mathcal{D}$ independently at random according to $\mu$.\footnote{So that $\mathbbm{P} \left[{\bf t}_j \in \mathcal{S} \right] = \mu \left( \mathcal{S} \right)$ for all measurable $\mathcal{S} \subseteq \mathcal{D}$ and $j \in [m]$.}  Then, form the matrix $R'$ by setting $R'_{i,j} := \psi_j \left(~ {\bf t}_i~\right)$ for each $i \in [m]$ and $j \in [N]$.  The following theorem concerning random sampling matrices created from bounded orthonormal systems is proven in \cite{HolgerBook}.\footnote{The specific form of the lower bound used for $m$ below is taken from Theorem 12.12 of \cite{HolgerBook}.}
\begin{thm}
Let $R' \in \mathbbm{R}^{m \times N}$ be a random sampling matrix created from a bounded orthonormal system with constant $K$.  Let $\mathcal{S} \subset [N]$ have have cardinality $|\mathcal{S}| = d$, and set $\widetilde{R'} = \frac{1}{\sqrt{m}}R'$.  Then, for $\epsilon \in (0,1)$, we will have
$$\sqrt{1 - \epsilon} ~\leq~ \sigma_d \left(\widetilde{R'}_{\mathcal{S}} \right)~\leq~\sigma_1\left( \widetilde{R'}_{\mathcal{S}} \right) ~\leq~ \sqrt{1 + \epsilon}$$
with probability at least $1-p$ provided that $m \geq (8/3)K^2 \epsilon^{-2} d \ln(2d / p)$.
\label{thm:randSamp}
\end{thm}
Note that Theorem~\ref{thm:randSamp} also applies to the special case where our orthonormal system, $\Psi$, consists of the $N$ columns of a rescaled unitary matrix $U \in \mathbbm{R}^{N \times N}$ (i.e., $\psi_j = \sqrt{N} {\bf U}_j$ for all $j \in [N]$).  Here, $\mathcal{D} = [N] \subset \mathbbm{R}$, $\psi_j(i) = \sqrt{N}U_{i,j}$ for all $i,j \in [N]$, and $\mu$ is the discrete uniform measure on $[N]$.  In this case we will consider the random sampling matrix, $R'$,  for $\Psi$ to be the product $\sqrt{N}RU$, where $R \in \{ 0,1\}^{m \times N}$ is a random matrix with exactly one nonzero entry per row (which is selected uniformly at random).  %We will occasionally abuse terminology by referring to any such random matrix, $R \in \{ 0,1\}^{m \times N}$, as a \textit{random sampling matrix}.
We will refer to any such random matrix, $R \in \{ 0,1\}^{m \times N}$, as a \textit{random selector matrix}.

%%%%%%%
%%%%%%%
\section{Exponential Accuracy for First Order Sigma-Delta via Random Sampling}
\label{sec:randSamp}

In this section we will deal only with first order Sigma-Delta.  Hence, given ${\bf x} \in \mathcal{B}^d$, the vectors ${\bf q}, {\bf u} \in \mathbbm{R}^N$ will always be those resulting from \eqref{equ:FirstOrdq} and \eqref{equ:FirstOrdu} above.  Our objective in this section is to demonstrate that a small set of sums of the bit stream produced by the first order scheme considered herein suffices to accurately encode the vector being quantized.  Furthermore, and somewhat surprisingly, the number of sums we must keep in order to successfully approximate the quantized vector is \textit{entirely independent of $N$} (though the reconstruction error depends on $N$).  Proving this will require the following lemma.

\begin{lem}
For every ${\bf x} \in \mathcal{B}^d$ we will have $D^{-1}{\bf q} \in \{ -N, \dots, N \}^N \subset \mathbbm{Z}^N$.
\label{lem:qInfBound}
\end{lem}

\noindent \textit{Proof:}  Note that $q_i \in \{ -1, 1 \}$ for all $i \in [N]$ (see \eqref{equ:FirstOrdq}).  Furthermore, it is not difficult to check that 
$$\left( D^{-1} \right)_{i,j} = \left\{ \begin{array}{ll} 1 & \textrm{if}~j \leq i\\ 0 & \textrm{otherwise} \end{array} \right..$$
Thus, we have that $\left( D^{-1}{\bf q} \right)_i \in \{ -i, \dots, i \}$ for all $i \in [N]$.~~$\Box$\\

We are new equipped to prove the main theorem of this section.

\begin{thm}
Let $\epsilon,p \in (0,1)$, and $R \in \{ 0,1\}^{m \times N}$ be a random %sampling
selector matrix.  Then, there exists a frame $F \in \mathbbm{R}^{N \times d}$ such that 
$$\left\| {\bf x} - \left(RD^{-1}F \right)^{\dagger} R D^{-1} {\bf q} \right\|_2 \leq \frac{ \sqrt{2} \pi}{\sqrt{1 - \epsilon}} \left( \frac{d^{\frac{3}{2}}}{N} \right)$$
for all ${\bf x} \in \mathcal{B}^d \subset \mathbbm{R}^d$ with probability at least $1-p$, provided that $m \geq (16/3) \epsilon^{-2} d \ln(2d / p)$. Here, $\mathbf q$ is the output of the first order $\sd$ quantization scheme  \eqref{equ:FirstOrdq} and \eqref{equ:FirstOrdu}, applied to $F\x$.  Furthermore, $R D^{-1} {\bf q}$ can always be encoded using $b \leq m ( \log_2 N + 1)$ bits.
\label{thm:ResExp}
\end{thm}

\noindent \textit{Proof:}  Let $U, \Sigma, V \in \mathbbm{R}^{N \times N}$ be defined as in \eqref{equ:LsingVecsD}, \eqref{equ:SingValsD}, and \eqref{equ:RsingVecsD}, respectively.  Define $F  \in \mathbbm{R}^{N \times d}$ to be the (renormalized) last $d$ columns of $U$,
\begin{equation}
F := \sqrt{\frac{N}{2d}} \left( {\bf U}_{N-d+1} \dots {\bf U}_N \right).
\label{equ:DefF}
\end{equation}
We refer to the frame corresponding to $F$ as the $1$st order Sobolev self-dual frame. 
Denoting the $i^{\rm th}$ row of $F$ by ${\bf f}_i \in \mathbbm{R}^d$, we note that \eqref{equ:LsingVecsD} implies that
\begin{equation}
\| {\bf y} \|_{\infty} ~=~ \left\| F {\bf x} \right\|_{\infty} ~\leq~ \max_{i \in [N]} \| {\bf f}_i \|_2 \| {\bf x} \|_2~\leq~\sqrt{\frac{N}{2d}}\sqrt{\frac{2d}{N+1/2}} \cdot \|{\bf x} \|_2 ~\leq~ 1
\label{equ:yLinfB}
\end{equation}
for all ${\bf x} \in \mathcal{B}^d$.  Now, apply the random %sampling 
selector matrix, $R$, to~\eqref{equ:Recon} to obtain
\begin{equation}
R{\bf u} ~=~ RD^{-1}F{\bf x} - RD^{-1}{\bf q}.\nonumber
\end{equation}
Since our goal is to obtain an upper bound on \begin{equation}  \|(RD^{-1}F)^\dagger R{\bf u}\|_2 ~=~ \| {\bf x} - (RD^{-1}F)^\dagger RD^{-1}{\bf q}\|_2 \label{equ:ErrorEst} \end{equation} and since $\|Ru\|_2$ is easily controlled (see the discussion after \eqref{equ:LinSigDelt}), it behooves us to study $RD^{-1}F \in \mathbbm{R}^{m \times d}$. Observe that 
\begin{equation}
D^{-1}F ~=~ \sqrt{\frac{N}{2d}} V \Sigma^{-1} U^{\rm T} \left( {\bf U}_{N-d+1} \dots {\bf U}_N \right) ~=~ \sqrt{\frac{N}{2d}} \left( ({\bf V})_{N-d+1} \dots ({\bf V})_N \right) ~\widetilde{\Sigma},
\label{equ:DinvF}
\end{equation}
where $\widetilde{\Sigma} \in \mathbbm{R}^{d \times d}$ has
$$\widetilde{\Sigma}_{i,j} ~=~ \frac{\delta_{i,j}}{\sigma_{N-d+j}(D)}.$$
Let $\mathcal{S} = \{ N-d+1, \dots, N \} \subset [N]$.  Then, 
$$RD^{-1}F ~=~ \sqrt{\frac{N}{2d}} R V_{\mathcal{S}} \widetilde{\Sigma} ~=~ \sqrt{\frac{m}{2d}} \left( \frac{\sqrt{N} R V}{\sqrt{m}} \right)_{\mathcal{S}} \widetilde{\Sigma}.$$
Applying Theorem~\ref{thm:randSamp} now tells us that 
\begin{equation}
\sqrt{\frac{m}{2d}} \cdot \frac{\sqrt{1 - \epsilon}}{\sigma_{N-d+1}(D)} ~\leq~ \sigma_d \left( RD^{-1}F \right)~\leq~\sigma_1\left( RD^{-1}F \right) ~\leq~ \sqrt{\frac{m}{2d}} \cdot \frac{\sqrt{1+ \epsilon}}{\sigma_{N}(D)}
\label{equ:wellCond}
\end{equation}
with probability at least $1-p$, provided that $m \geq (16/3) \epsilon^{-2} d \ln(2d / p)$.

Whenever \eqref{equ:wellCond} holds we may approximate ${\bf x} \in \mathcal{B}^d$ by 
$$\hat{\bf x} : = \left(RD^{-1}F \right)^{\dagger} R D^{-1} {\bf q},$$
and then use \eqref{equ:ErrorEst} to estimate the approximation error as
$$\| {\bf x} - \hat{\bf x} \|_2 ~=~ %\left\| {\bf x} - \left(RD^{-1}F \right)^{\dagger} R D^{-1} {\bf q} \right\|_2 ~=~ 
 \left\| \left(RD^{-1}F \right)^{\dagger} R {\bf u} \right\|_2 ~\leq~ \sqrt{\frac{2d}{m}} \cdot \frac{\sigma_{N-d+1}(D)}{\sqrt{1 - \epsilon}} \left\| R {\bf u} \right\|_2.$$
Using \eqref{equ:SingValsD} and recalling that $|u_i| \leq 1$ for all $i \in [N]$ since $|y_i| \leq 1$ for all $i \in [N]$ (see \eqref{equ:yLinfB}), we obtain 
$$\| {\bf x} - \hat{\bf x} \|_2 ~\leq~ \frac{2 \sqrt{2d}}{\sqrt{1 - \epsilon}} \cdot \cos \left( \frac{(N-d+1) \pi}{2N + 1} \right) ~\leq~ \frac{2 \sqrt{2d}}{\sqrt{1 - \epsilon}} \cdot \left( \frac{\pi}{2} - \frac{(N-d+1) \pi}{2N + 1}\right) ~\leq~\frac{\sqrt{2}\pi}{\sqrt{1 - \epsilon}} \left( \frac{d^{\frac{3}{2}}}{N} \right).$$
Finally, Lemma~\ref{lem:qInfBound} tells us that $R D^{-1} {\bf q}$ can always be encoded using $b \leq m ( \log_2 N + 1)$ bits.~~$\Box$\\

\begin{rem}
Theorem~\ref{thm:ResExp} provides the desired exponentially decaying rate-distortion bounds.  In particular, by choosing the smallest integer $m \geq (16/3) \epsilon^{-2} d \ln(2d / p)$, the rate is 
$$\mathcal{R} = m ( \log_2 N + 1),$$ 
and the distortion is 
$$\mathcal{D} = \frac{\sqrt{2} \pi d^{3/2}}{N\sqrt{1 - \epsilon}}.$$  
Expressing the distortion in terms of the rate, we obtain 
\begin{equation}
\mathcal{D}(\mathcal{R}) = \frac{2 \sqrt{2} \pi d^{3/2}}{\sqrt{1 - \epsilon}} \cdot 2^{-\mathcal{R}/m} = C_1(\epsilon) \cdot d^{3/2} \exp \left( - \frac{\mathcal{R}}{C_2(\epsilon) d \ln(2d/p)} \right).
\label{equ:RateDist1}
\end{equation}
Above, $C_1(\epsilon) =  2 \pi \cdot \sqrt{\frac{2}{1-\epsilon}}$ and $\frac{16}{3  \ln 2 \cdot \epsilon^2} + \frac{1}{d \ln 2 \cdot \ln(2d/p)} \geq C_2(\epsilon) \geq \frac{16}{3  \ln 2 \cdot \epsilon^2}$.
\end{rem}

Theorem~\ref{thm:ResExp} demonstrates that only $O(\log N)$-bits must be saved and/or transmitted in order to achieve $O(1/N)$-accuracy (neglecting other dependencies). Moreover, from a practical point of view the use of random sampling matrices is appealing as they allow for the $O(\log N)$ bits to be computed ``on the fly" and with minimal cost. However, the first order scheme we consider herein has deficiencies that merit additional consideration.  Primarily, the first order scheme \eqref{equ:FirstOrdq} and \eqref{equ:FirstOrdu} must still be executed before its output bitstream can be compressed.  Hence, utilizing the compressed sigma delta encoding described in this section has an $O(N)$ ``cost"  associated with it (as the $N$-dimensional vector $F\x$ must be acquired and quantized). This cost may be prohibitive for some applications.  Hence, we will consider results for higher order schemes (and more general frames) in the next section.

%%%%%%%
%%%%%%%
\section{Exponential Accuracy For General Frames and Orders via Bernoulli Random Matrices}
\label{sec:BernSamp}
In this section we will deal with a more general class of stable $r^{\rm th}$-order Sigma-Delta schemes.  Hence, given ${\bf x} \in \mathcal{B}^d$, the vectors ${\bf q}, {\bf u} \in \mathbbm{R}^N$ will always be those resulting from \eqref{equ:rthOrdq} and \eqref{equ:rthOrdu} above.  The main result of this section will require the following lemma, which is essentially proven in \cite{baraniuk2008simple}.

\begin{lem}
Let $\epsilon,p \in (0,1)$, $\{ {\bf v}_1, \dots, {\bf v}_d \} \subset \mathbbm{R}^N$, and $B \in \{ -1, 1\}^{m \times N}$ be a Bernoulli random matrix.  Set $\widetilde{B} = \frac{1}{\sqrt{m}}B$.  Then, 
$$(1 - \epsilon) \| ~{\bf x }~ \|_2 \leq \|~ \widetilde{B} {\bf x}~ \|_2 \leq (1 + \epsilon) \|~ {\bf x}~ \|_2$$
for all ${\bf x} \in \textrm{span} \{ {\bf v}_1, \dots, {\bf v}_d \}$ with probability at least $1 - p$, provided that
$$m \geq \frac{4 d \ln (12/ \epsilon) + 2 \ln(1/p)}{\epsilon^2/8 - \epsilon^3/24}.$$
\label{lem:JLsubspace}
\end{lem}

\noindent \textit{Proof:}  Combine Theorem~\ref{thm:BernJL} with the proof of Lemma 5.1 and the subsequent discussion in \cite{baraniuk2008simple}.~~$\Box$\\

In addition to considering more general $r^{\rm th}$-order quantization schemes, we will also consider a more general class of frames, $F  \in \mathbbm{R}^{N \times d}$.  More specifically, we will allow any frame matrix which adheres to the following definition.
\begin{Def}
We will call a frame matrix $F \in \mathbbm{R}^{N \times d}$ an $(r,C,\alpha)$-frame if
\begin{enumerate}
\item  $\| F{\bf x} \|_{\infty} \leq 1$ for all ${\bf x} \in \mathcal{B}^d$, and
\item $\sigma_d \left( D^{-r} F \right) \geq C \cdot N^{\alpha}$.
\end{enumerate}
\label{def:craframe}
\end{Def}
Roughly speaking, the first condition of Definition~\ref{def:craframe} ensures that the frame $F$ is uniformly bounded, while the second condition can be interpreted as a type of smoothness requirement.  We are now properly equipped to prove the main theorem of this section.  

\begin{thm}
Let $\epsilon,p \in (0,1)$, $B \in \{ -1, 1\}^{m \times N}$ be a Bernoulli random matrix, and $F \in \mathbbm{R}^{N \times d}$ be an $(r,C,\alpha)$-frame with $r \in \mathbbm{N}$, $\alpha \in (1,\infty)$, and $C \in \mathbbm{R}^+$.  Consider $\mathbf{q}$, the quantization of $F\x$ via a stable $r^{th}$-order scheme with alphabet $\mathcal{A}_A^{2\mu}$ and stability constant ${C}_{\rho,Q}(r) \in \mathbbm{R}^+$  (see \eqref{equ:rthOrdq}, \eqref{equ:rthOrdu}, \eqref{eq:midrise} and the subsequent discussion).  Then,  the following are true.
\vspace{ +5pt} \begin{enumerate}\itemsep +10pt
\item[(i)] The reconstruction error (i.e., the distortion) satisfies 
$$\left\| {\bf x} - \left({B}D^{-r}F \right)^{\dagger} {B} D^{-r} {\bf q} \right\|_2 ~\leq~ \frac{{C}_{\rho,Q}(r) \cdot N^{1 - \alpha}}{C \cdot (1 - \epsilon)}$$
for all ${\bf x} \in \mathcal{B}^d \subset \mathbbm{R}^d$ with probability at least $1-p$, provided that $m \geq \frac{4 d \ln (12/ \epsilon) + 2 \ln(1/p)}{\epsilon^2/8 - \epsilon^3/24}$. 
\item[(ii)] $B D^{-r} {\bf q}$ can always be encoded using $b \leq m [ (r+1) \log_2 N + \log_2 A + 1]$ bits.
\end{enumerate}
\label{thm:BerExp}
\end{thm}

\noindent \textit{Proof:}  Apply a Bernoulli random matrix, $B \in \{ -1, 1 \}^{m \times N}$, to~\eqref{equ:ReconOrdr} and then renormalize by $m^{-1/2}$ to obtain 
\begin{equation}
\widetilde{B}{\bf u} ~=~ \widetilde{B}D^{-r}F{\bf x} - \widetilde{B}D^{-r}{\bf q},
\label{equ:ErrorEst2}
\end{equation}
where $\widetilde{B} = \frac{1}{\sqrt{m}}B$. Considering $\widetilde{B}D^{-r}F \in \mathbbm{R}^{m \times d}$, we note that Lemma~\ref{lem:JLsubspace} guarantees that $\widetilde{B}$ is an near-isometry on $\textrm{span} \{ D^{-r}{\bf F}_1, \dots, D^{-r}{\bf F}_d \}$.  Thus,
\begin{equation}
(1 - \epsilon) \cdot C \cdot N^{\alpha} ~\leq~ \sigma_d \left( \widetilde{B}D^{-r}F \right)
 \label{equ:wellCond2}
\end{equation}
with probability at least $1-p$, provided that $m \geq \frac{4 d \ln (12/ \epsilon) + 2 \ln(1/p)}{\epsilon^2/8 - \epsilon^3/24}$.  

Given that \eqref{equ:wellCond2} holds, we may approximate ${\bf x} \in \mathcal{B}^d$ using $B D^{-r} {\bf q} \in \mu \mathbbm{Z}^m$ by 
$$\hat{\bf x} : = \frac{1}{\sqrt{m}} \left(\widetilde{B}D^{-r}F \right)^{\dagger} B D^{-r} {\bf q} ~=~\left(\widetilde{B}D^{-r}F \right)^{\dagger} \widetilde{B} D^{-r} {\bf q},$$
and then use \eqref{equ:ErrorEst2} to estimate the approximation error as
$$\| {\bf x} - \hat{\bf x} \|_2 ~=~ \left\| \left(\widetilde{B}D^{-r}F \right)^{\dagger} \widetilde{B} {\bf u} \right\|_2 ~=~ \left\| \left({B}D^{-r}F \right)^{\dagger} {B} {\bf u} \right\|_2 ~\leq~ \frac{N^{- \alpha}}{C \cdot (1 - \epsilon)} \left\| \widetilde{B} {\bf u} \right\|_2.$$
Noting that $\| {\bf u} \|_{\infty} \leq {C}_{\rho,Q}(r)$ since $\| F{\bf x} \|_{\infty} \leq 1$ (by definition of $(r,C,\alpha)$-frames), we obtain 
$$\| {\bf x} - \hat{\bf x} \|_2 ~\leq~ \frac{N^{- \alpha}}{C \cdot (1 - \epsilon)} \left\| \widetilde{B} {\bf u} \right\|_2 ~\leq~\frac{{C}_{\rho,Q}(r) \cdot N^{1 - \alpha}}{C \cdot (1 - \epsilon)}.$$
Finally, a short argument along the lines of Lemma~\ref{lem:qInfBound} tells us that $B D^{-r} {\bf q} \in \mu \mathbbm{Z}^m$ will always have $\| B D^{-r} {\bf q} \|_{\infty} \leq 2 \mu A \cdot N^{r+1}$.  Thus, $B D^{-r} {\bf q}$ can be encoded using $b \leq m [ (r+1) \log_2 N + \log_2 A + 1]$ bits.  Note that $\mu$ does not influence the number of required bits.~~$\Box$\\

\begin{rem}
Theorem~\ref{thm:BerExp} also provides the desired exponentially decaying rate-distortion bounds.  In particular, by choosing the smallest integer $m \geq \frac{4 d \ln (12/ \epsilon) + 2 \ln(1/p)}{\epsilon^2/8 - \epsilon^3/24}$, the rate is 
$$\mathcal{R} = m [ (r+1) \log_2 N + \log_2 A + 1],$$ 
and the distortion is 
$$\mathcal{D} = \frac{{C}_{\rho,Q}(r) \cdot N^{1 - \alpha}}{C \cdot (1 - \epsilon)}.$$  
Expressing the distortion in terms of the rate, we obtain 
\begin{equation}
\mathcal{D}(\mathcal{R}) = \frac{{C}_{\rho,Q}(r) \cdot (2A)^{(\alpha-1)/(r+1)}}{C \cdot (1 - \epsilon)} \cdot 2^{-(\mathcal{R} (\alpha - 1)/m(r+1))} ~\leq~ \bar{C}_{\rho,Q}(A,\epsilon,\alpha,r) \cdot \exp \left( - \frac{\mathcal{R}}{d \cdot C_3(\epsilon,p)} \right).
\label{equ:RateDist2}
\end{equation}
Above, $\bar{C}_{\rho,Q}(A,\epsilon,\alpha,r) =  \frac{{C}_{\rho,Q}(r) \cdot (2A)^{(\alpha-1)/(r+1)}}{C \cdot (1 - \epsilon)}$ and $\frac{4 (r+1)\ln \left( 12 / \epsilon p \right)}{\ln 2 \cdot (\alpha-1) (\epsilon^2/8 - \epsilon^3/24)} + \frac{1}{d} \geq C_3(\epsilon,p) > 0$. 
%\geq \frac{4 (r+1)\ln \left( 12 / \left(\epsilon \cdot p^{1/2d} \right) \right)}{\ln 2 \cdot (\alpha-1) (\epsilon^2/8 - \epsilon^3/24)}$.
\end{rem}

\begin{rem}
The choice of Bernoulli matrices in Theorem~\ref{thm:BerExp} is motivated by two properties. First,  one can encode their action on the integrated $\sd$ bit-stream in a lossless manner. Second, Bernoulli matrices (of appropriate size) act as near isometries on $\textrm{span} \{ D^{-r}{\bf F}_1, \dots, D^{-r}{\bf F}_d \}$. In fact, any encoding matrix drawn from a distribution satisfying the above two properties would work for compressing the $\sd$ bit-stream. For example, \cite{achlioptas2001} also studied other discrete random matrices (whose entries are $\pm 1$ with probability $1/6$ each, and $0$ with probability $2/3$) and showed that they serve as Johnson-Lindenstrauss embeddings. %In some cases, one may be interested in fast computation and may thus use a fast discrete Johnson-Lindenstrauss embedding.  For example, matrices of the form $B=RHS$ where $R$ is a random sampling operator, $H$ is a Hadamard matrix and $S$, is  a diagonal matrix with equiprobable $\pm 1$ entries on its diagonal (cf. \cite{krahmer2010new}) can be applied in time proportional to . Such an $m\times N$ matrix must satisfy $m\gtrsim \frac{1}{\epsilon^2}d\log^3{d}\log{N}$ to act as a near isometry in our subspace of interest [I THINK THIS MIGHT MEAN SUB-EXPONENTIAL]. 
\end{rem}

It is informative to compare the rate-distortion bounds resulting from Theorems~\ref{thm:ResExp} and~\ref{thm:BerExp} in the case of the first order Sigma-Delta scheme (and frame) considered by Theorem~\ref{thm:ResExp}.  If we define $F \in \mathbbm{R}^{N \times d}$ as per \eqref{equ:DefF} we can see, by considering \eqref{equ:yLinfB} and \eqref{equ:DinvF}, that it will be a $\left(1, d^{-3/2}(\sqrt{2} \pi)^{-1}, 3/2 \right)$-frame.  Furthermore, the first order scheme considered by Theorem~\ref{thm:ResExp} has $A = 1$ and ${C}_{\rho,Q}(1) = 1$.  Hence, we see that \eqref{equ:RateDist2} becomes
$$\mathcal{D}(\mathcal{R}) = \frac{2^{3/4} \pi d^{3/2}}{1 - \epsilon} \cdot 2^{-\mathcal{R} / 4m} ~\leq~ \frac{2^{3/4} \pi d^{3/2}}{1 - \epsilon} \cdot 2^{-\mathcal{R} \left( \frac{\epsilon^2 - \epsilon^3 / 3}{128 \cdot d \ln (12/ \epsilon) + 64 \ln(1/p)} \right)} $$
in this case.  Comparing this expression to \eqref{equ:RateDist1} we can see that the dependence on $d$ has been improved (i.e., by a log factor) in the denominator of the exponent.  However, we have sacrificed some computational simplicity for this improvement since $B D^{-1} {\bf q}$ will generally require more effort to compute then $RD^{-1}{\bf q}$ in practice.

Importantly, though, Theorem~\ref{thm:BerExp} also enables one to obtain near-optimal rate-distortion bounds. Moreover, fixing the desired distortion, fewer samples $N$ may now be used than in Theorem~\ref{thm:ResExp} (hence less computation for quantization and encoding) via the use of higher order quantization schemes.  As a result, we will be able to use $(r,C,\alpha)$-frames having only $O({N}^{\frac{1}{\alpha-1}})$ rows below while still achieving $O(1/N)$ accuracy (ignoring dependences on other parameters such as $d$, etc.).  This represents a clear improvement over the first order scheme we have considered so far for all $\alpha>2$, provided such $(r,C,\alpha)$ frames exists.
%Thus, we must obtain more general $(r,C,\alpha)$-frames for $r \geq2$ before higher order schemes can be utilized. %in conjunction with Theorem~\ref{thm:BerExp}.  
In the next section we will briefly survey some examples of currently known $(r,C,\alpha)$-frames, for general $r \in \mathbbm{N}$, which are suitable for use with the type of stable $r^{\rm th}$-order sigma delta schemes considered herein.

\subsection{Examples of $(r,C,\alpha)$-frames}

In this section we briefly survey some $(r,C,\alpha)$-frames that can by utilized in concert with Theorem~\ref{thm:BerExp} above.

\begin{ex}{\bf Sobolev self-dual frames \cite{krahmer2012root}}\label{ex:sobDual}\end{ex}
Our first example of a family of $(r,C,\alpha)$-frames represents a generalization of the frame utilized by Theorem~\ref{thm:ResExp} to higher orders. Let $U_{D^r} = \left( \mathbf{U}_1 ... \mathbf{U}_N\right)$ be the matrix of left singular vectors of $D^r$, corresponding to a decreasing arrangement of the singular values. Then, we refer to $F_{(r)}=\left( \mathbf{U}_{N-d+1} ... \mathbf{U}_N\right)$ as the ($r^{\rm th}$-order) Sobolev self-dual frame. $F_{(r)}$ is an $(r,C,\alpha)$-frame with $C= \pi^{-r} ( d+2r )^{-r}$ and $\alpha = r$ (see \cite{krahmer2012root}, Theorem 8).

For these frames, with fixed $r$, using the $\sd$ schemes of Proposition \ref{prop:best_known} and a Bernoulli encoding matrix, the exponent in the rate-distortion expression $\mathcal{D}(\mathcal{R})$ behaves like $-\frac{r-1}{r+1}\frac{\mathcal{R}}{d}$. Specifically, considering Example~\ref{ex:sobDual} with $r=1$ we see that \eqref{equ:DefF}, when unnormalized, is a $(1, \pi^{-1} (d+2)^{-1}, 1)$-frame instead of a $(1, d^{-3/2}(\sqrt{2} \pi)^{-1}, 3/2 )$-frame.  Hence, the bound provided for the unscaled matrix by Example~\ref{ex:sobDual} with $r=1$ is weaker  than the result for the rescaled matrix, whenever $d$ is significantly smaller than $N$.  %In general, rescaling The lesson here is that rescaling $F$ so that $\| F {\bf x} \|_{\infty}$ is as near $1$ as possible for all ${\bf x} \in \mathcal{B}^d$ is generally a good idea.  
What prevents us from rescaling $F$ when $r\geq 2$ is that we have insufficient information regarding the left singular vectors of $D^r$.   %However, in practice one might attempt to capitalize on this observation by $(i)$ rescaling $F {\bf x}$ by a large constant (e.g., by a power of 2) on the fly before quantization whenever $\| F {\bf x} \|_{\infty}$ is small, and then $(ii)$ transmitting the rescaling constant's setting along with the quantized bit stream for the purpose of recovery.\\
\begin{ex}{\bf Harmonic frames}\label{ex:harm}\end{ex}
A \textit{harmonic frame}, $F \in \mathbbm{R}^{N \times d}$, is defined via the following related functions:
\begin{equation}
F_0(t) ~=~ \frac{1}{\sqrt{2}},
\end{equation}
\begin{equation}
F_{2j-1}(t) ~=~ \cos(2 \pi j t),~~j \geq 1,~~\textrm{and}
\end{equation}
\begin{equation}
F_{2j}(t) ~=~\sin(2 \pi j t),~~j \geq 1.
\end{equation}
We then define $F_{j,k} = \sqrt{\frac{2}{d}} \cdot F_{j'}(k/N)$, where $j' = j-d~\textrm{mod}~2$ for all $k \in [N]$ and $j \in [d+d~\textrm{mod}~2]$.  In addition to Sobolev self-dual frames, we note that harmonic frames also yield general $(r, C, \alpha)$-frames.  For sufficiently large $N$, a harmonic frame is an $(r,C,\alpha)$-frame with $C= C_1 e^{r/2} r^{-(r+C_2)}$ and $\alpha = r+1/2$ (see \cite{krahmer2012root}, Lemma 17). Here, $C_1$ and $C_2$ are constants that possibly depend on $d$. For this example, with fixed $r$, using the $\sd$ schemes of Proposition \ref{prop:best_known} and a Bernoulli encoding matrix, the exponent in the rate-distortion expression $\mathcal{D}(\mathcal{R})$ behaves like $-\frac{r-1/2}{r+1}\frac{\mathcal{R}}{d}$.

\begin{ex}{\bf Frames generated from piecewise-$C^1$ uniformly sampled frame paths}\label{ex:smooth}
\end{ex}

Note that the example above is a  special case of  a \textit{smooth frame} \cite{blum:sdf}.  As one might expect, more general classes of smooth frames also yield $(r,C,\alpha)$-frames. One such class of frames consists of those generated from piecewise-$C^1$ uniformly sampled frame paths, as defined in \cite{blum:sdf}. For convenience, we reproduce the definition below. 
\begin{Def}
A vector valued function $E: [0,1]\mapsto \R^d$ given by $E(t)=[\mathbf{E}_1(t), \mathbf{E}_2(t),...,\mathbf{E}_d(t)]$ is a piecewise-$C^1$ uniformly sampled frame path if 
\begin{enumerate}
\item for all $n\in [d]$, $\mathbf{E}_n: [0,1] \mapsto \mathbbm{R}$ is piecewise-$C^1$,
\item the functions $\mathbf{E}_n$ are linearly independent, and
\item there exists an $N_0$ such that for all $N\geq N_0$, the matrix $F$ with entries $F_{ij} = \mathbf{E}_j(i/N)$ is a frame matrix. 
\end{enumerate}
In this case, we say that the frame $F$ is generated from a piecewise-$C^1$ uniformly sampled frame path. 
\end{Def}
For any piecewise-$C^1$ uniformly sampled frame path, there is an $N_0 \in \mathbb N$ such that for all $N>N_0$, any frame generated from the frame path is an $(r,C,\alpha)$-frame for some $C$ (possibly depending on $r$ and $d$) and $\alpha = r+1/2$ (see \cite{blum:sdf}, Theorem 5.4 and its proof). Here, again, with fixed $r$, using the $\sd$ schemes of Proposition \ref{prop:best_known} and a Bernoulli encoding matrix, the exponent in the rate-distortion expression $\mathcal{D}(\mathcal{R})$ behaves like $-\frac{r-1/2}{r+1}\frac{\mathcal{R}}{d}$. Example~\ref{ex:smooth} deals with smooth frames of a fairly general type, albeit at the cost of less precision in specifying $C$.  Perhaps more surprisingly, decidedly non-smooth frames also yield $(r,C,\alpha)$-frames in general.  In particular, we may utilize Bernoulli random matrices as both our bit stream compression operator, \textit{and} our $(r,C,\alpha)$-frame.

\begin{ex}{\bf Bernoulli and Sub-Gaussian frames}\label{ex:Bernoulli}\end{ex}
Let $\gamma \in [0,1]$. Then, there exists constant $c_1$ and $c_2$, such that with probability exceeding $1-2e^{-c_1 N^{1-\gamma}d^\gamma/4}$, a frame $F$ whose entries are $\pm \frac{1}{\sqrt{d}}$ Bernoulli random variables is an $(r,C,\alpha)$-frame, provided $N\geq (c_2 r)^\frac{1}{1-\gamma} d$. Here $C=d^{-\gamma(r-1/2)+1/2}$ and $\alpha=1/2 + \gamma(r-1/2)$. See \cite[Proposition 14]{KSY12} for a proof. 

In fact, Bernoulli frames are a special case of a more general class of frames whose entries are sub-Gaussian random variables.  These more general types of random matrices also serve as $(r,C,\alpha)$-frames.

%\begin{ex}{\bf Sub-Gaussian frames}\label{ex:subG}\end{ex}
%We will need the following definitions.  \begin{Def}A zero-mean random variable $\eta$ is called sub-Gaussian if there exists a constant $c>0$ such that $\mathbb{E} e^{t\eta} \leq e^{c^2 t^2/2}$ for all $t\in\R$.\label{def:SG}\end{Def}
%
%\begin{Def} If a sub-Gaussian random variable $\eta$ satisfies $P(|\eta|>t) \leq K P(|\xi|>t)$ for all $t\geq0$,  where $\xi$ is a Gaussian random variable drawn according to $\mathcal{N}(0,\sigma^2)$, then we say that $\eta$ is $K$-dominated by $\mathcal{N}(0,\sigma^2)$. \end{Def}
\begin{Def} If two random variables $\eta$ and $\xi$ satisfy  $P(|\eta|>t) \leq K P(|\xi|>t)$ for some constant $K$ and all $t\geq0$ then we say that $\eta$ is $K$-dominated by $\xi$. \end{Def}

%\begin{Def} We say that a random variable is sub-Gaussian with parameter $c>0$ if it is $e$-dominated by a Gaussian random variable with mean $0$ and variance $c^2$. \end{Def}

\begin{Def} We say that a matrix is sub-Gaussian with parameter $c$, mean $\mu$, and variance $\sigma^2$ if its entries are independent and $e$-dominated by a Gaussian random variable with parameter c, mean $\mu$, variance $\sigma^2$. \end{Def}

Let $\gamma \in [0,1]$. Then, there exists a constant $c_1>0$ such that, with probability exceeding $1-3e^{-c_1 N^{1-\gamma}d^\gamma}$, a random sub-Gaussian frame matrix $F$ with mean zero, variance $1/N$ , and parameter $c$ will be a $(r,C,\alpha)$-frame whenever $\frac{N}{d}\geq (c_2 r)^\frac{1}{1-\gamma}$ where $c_2$ depends only on $c$. Here $C=d^{-\gamma(r-1/2)}$ and $\alpha= \gamma(r-1/2)$. See \cite[Propositions 14 and 15]{KSY12} for a proof.  Consequently, using the $\sd$ schemes of Proposition \ref{prop:best_known} together with a Bernoulli encoding matrix and a Sub-Gaussian frame results in the exponent of the rate-distortion expression, $\mathcal{D}(\mathcal{R})$, behaving like $-\frac{\gamma \cdot r-\frac{1}{2}(\gamma+2)}{r+1}\frac{\mathcal{R}}{d}$.

%Let $\gamma \in [0,1]$. Then, there exists a constant $c>0$ such that, with probability exceeding $1-2e^{-cN^{1-\gamma}d^\gamma}$, a random frame $F$ whose entries are all $K$-dominated by $\mathcal{N}(0,1/N)$ will be a $(r,C,\alpha)$-frame whenever $N\geq d \Big(4\log\big((18\pi r^2)^3 N^2\big)\Big)^\frac{1}{1-\gamma}$. Here $C=\frac{e}{2}(3\pi r)^{-r} d^{-\gamma(r-1/2)}$ and $\alpha= \gamma(r-1/2)$. See \cite{KSY12} for a proof.  Consequently, using the $\sd$ schemes of Proposition \ref{prop:best_known} together with a Bernoulli encoding matrix and a Sub-Gaussian frame results in the exponent of the rate-distortion expression, $\mathcal{D}(\mathcal{R})$, behaving like $-\frac{\gamma \cdot r-\frac{1}{2}(\gamma+2)}{r+1}\frac{\mathcal{R}}{d}$. 

%%%%%%%
%%%%%%%
\section{Numerical experiments}
\label{sec:Numerical}
\begin{figure}[t]\
\centering
  \includegraphics[width=1\textwidth]{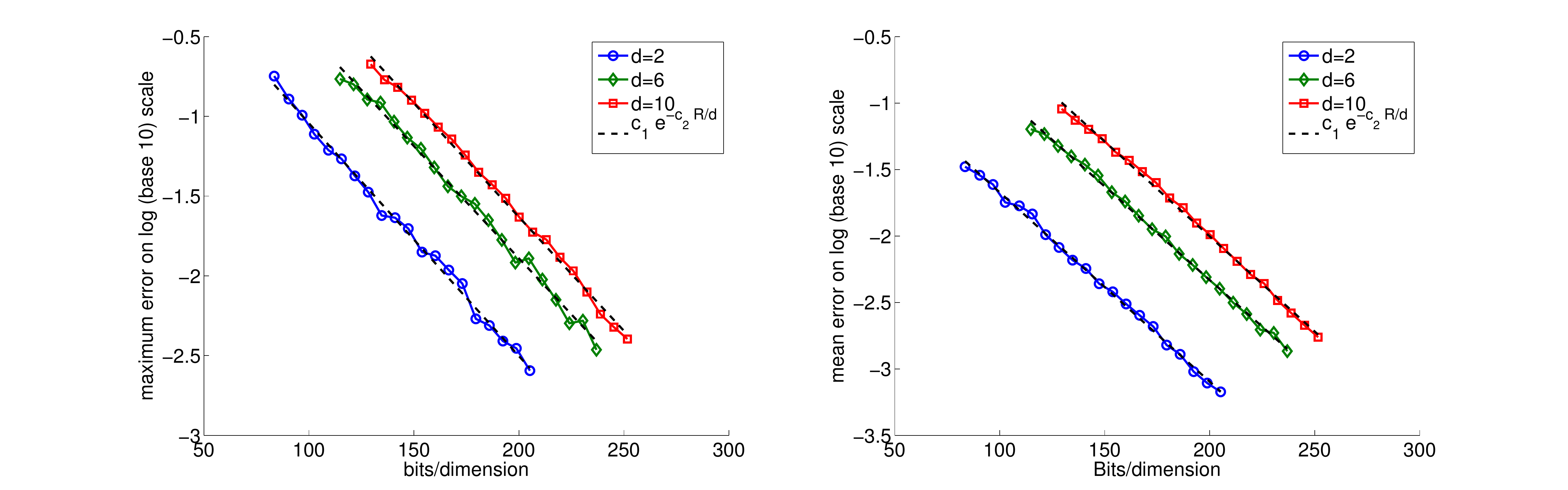}
\caption{(left) The maximum and (right) mean $\ell_2$-norm error (in $\log_{10}$ scale) plotted against the number of bits per dimension ($b/d$). Here we use a 1st order greedy $\sd$ scheme to quantize and a random %sampling 
selector matrix  to encode. \label{fig:SSD_frame_Random}}
\end{figure}

In this section we present numerical experiments to illustrate our results. To illustrate the results of Section \ref{sec:randSamp}, we first generate 5000 points uniformly from $\mathcal{B}^d$, with $d=2, 6, \text{and } 10$. We then compute, for various $N$, the $1$-bit $1$st order greedy $\sd$-quantization of $F\x$, where $F$ is an $N \times d$  Sobolev self-dual frame. $RD^{-1}\q$, where $R$ is an ${m\times N}$ random selector %sampling 
matrix with $m=10d$ is then employed to recover an estimate $\hat{\x}= (RD^{-1}F)^\dagger RD^{-1}\q$ of $\x$. In Figure \ref{fig:SSD_frame_Random} we plot (in log scale) the maximum and mean of  $\|\hat{\x}-\x\|_2$ over the 5000 realizations of $x$ versus the induced bit-rate. 

Our second experiment is similar, albeit we now use a third order 1-bit $\sd$ quantizer according to the schemes of \cite{deift2011optimal} to quantize the harmonic frame expansion of vectors in $\mathcal{B}^d$, with $d=2, 6, \text{and } 10$. Here, we use a $d\times m$ Bernoulli matrix, with $m=5d$ to encode $BD^{-1}\q$ and subsequently obtain $\hat{\x}= (BD^{-1}F)^\dagger BD^{-1}\q$. As before, we plot the maximum and mean of $\|\hat{\x}-\x\|_2$ over the 5000 realizations of $\x$ versus the induced bit-rate.

\begin{figure}[t]\
\centering
  \includegraphics[width=1\textwidth]{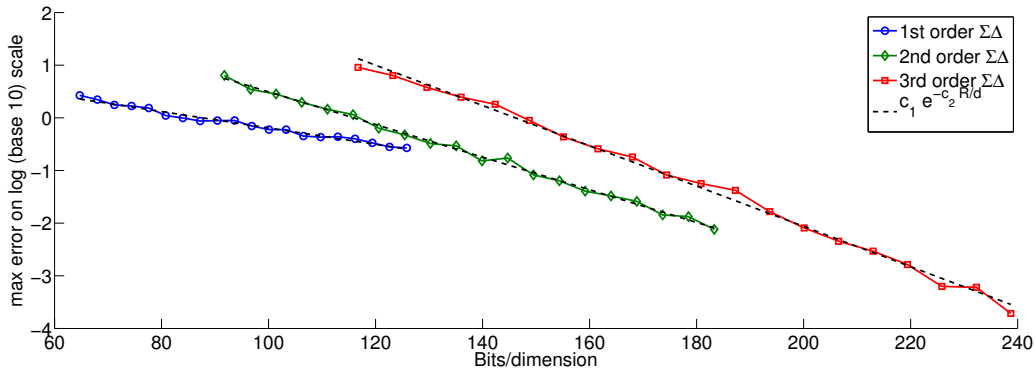}
\caption{ The maximum $\ell_2$-norm error (in $\log_{10}$ scale) plotted against the number of bits per dimension ($b/d$). Here $d=20$ and $\sd$ schemes with $r=1,2$ and $3$  are used to quantize the frame coefficients. A Bernoulli matrix is used for encoding. \label{fig:SSD_frame_Bernoulli2}}
\end{figure}

For our third experiment, we fix  $d=20$ and use the $\sd$ schemes of \cite{deift2011optimal} with $r=1,2, \text{ and } 3$ to quantize the Bernoulli frame coefficients, and we use Bernoulli matrices with $m=5d$ to encode. In Figure \ref{fig:SSD_frame_Bernoulli2} we show the maximum error versus the bit-rate. Note the different slopes corresponding to $r=1,2, \text{and } 3$. This observation is in agreement with the prediction (see the discussion around Example \ref{ex:Bernoulli}) that the exponent in the rate-distortion expression $\mathcal{D}(\mathcal{R})$ is a function of $r$. 
\begin{figure}[t]\
\centering
  \includegraphics[width=1\textwidth]{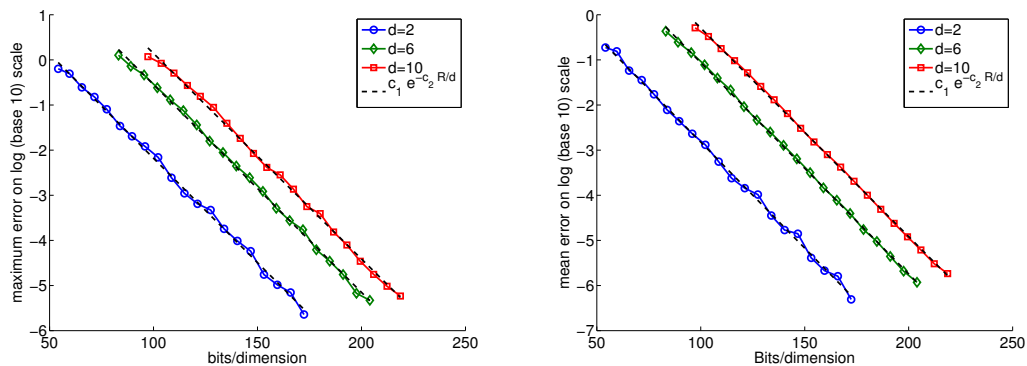}
\caption{(left) The maximum and (right) mean $\ell_2$-norm error (in $\log_{10}$ scale) plotted against the number of bits per dimension ($b/d$). Here we use a third order  $\sd$ scheme to quantize and a Bernoulli matrix  to encode. \label{fig:SSD_frame_Bernoulli}}
\end{figure}

%
%%%%%%%
%%%%%%%
%\section{things to add}
%\begin{itemize}
%\item talk about robustness being preserved. 
%\item talk about the Universality of Bernoulli frames, i.e., whatever the $(r,C,\alpha)$-frame is, Bernoulli will work - (maybe in a remark, maybe add it in the contributions?)  (DONE?).
%\item talk about how we are not married to Bernoulli, any kind of discrete concentration of measure type thing will work - e.g., fast things (DONE?). 
%\item repetition frame example + yack about bandlimited decimation 
%\item reference to Ulas's thesis. 
%\item comment about why section 3 is "worse" and "better" than section 4 (DONE?). 
%\item A remark after the Bernoulli theorem about how it enables one to obtain exponential accuracy with fewer samples than with first order sigma-delta by using higher order schemes (DONE?). 
%\item A section with examples: piecewise smooth frames (DONE?), random frames (e.g., Bernoulli), Sinan/Felix's SD schemes (DONE?).  
%\item Maybe (???) comment about how this is essentially vector quantization without the need for storing a codebook (or a data structure) (???). --- think about this some more. 
%\item Numerical experiments (DONE?). 
%\end{itemize}

\bibliographystyle{abbrv}
\bibliography{Iwen_Saab}

\end{document}